\documentclass[aps,10pt,prl,reprint,twocolumn,floatfix,showpacs,superscriptaddress]{revtex4-2}  

\usepackage{amsfonts,amsmath,amssymb}
\usepackage{comment}
\usepackage[margin=20mm,inner=20mm,marginpar=10mm]{geometry}
\usepackage{graphicx, epstopdf}
\usepackage{microtype}
\usepackage[figuresright]{rotating}
\usepackage{setspace}
\usepackage{textcomp}
\usepackage{times}
\usepackage{amsmath}
\usepackage{bm}
\usepackage[normalem]{ulem}
\usepackage[abs]{overpic}
\usepackage{color} 
\usepackage[dvipsnames]{xcolor}
\usepackage[T1]{fontenc}
\usepackage{url}
\usepackage[normalem]{ulem}

\newlength{\figwidth}
\newcommand{\angstrom}{\mbox{\normalfont\AA}}

\begin{document}

\title{
Atom diffraction in the strong-coupling regime
}

\author{Carina Kanitz}
\affiliation{German Aerospace Center (DLR), Institute of Quantum Technologies, Wilhelm-Runge-Stra\ss e 10, 89081 Ulm, Germany}

\author{Jakob Bühler}
\affiliation{German Aerospace Center (DLR), Institute of Quantum Technologies, Wilhelm-Runge-Stra\ss e 10, 89081 Ulm, Germany}

\author{Fran\c{c}ois Aguillon}
\affiliation{Institut des Sciences Mol\'eculaires d'Orsay (ISMO), Centre national de la
recherche scientifique (CNRS), University Paris-Sud, Universit\'e Paris-Saclay, Orsay, France}

\author{Vladimír Zobač}
\affiliation{University of Vienna, Faculty of Physics, Boltzmanngasse 5, 1090 Vienna, Austria}

\author{Jaime Glerum}
\affiliation{German Aerospace Center (DLR), Institute of Quantum Technologies, Wilhelm-Runge-Stra\ss e 10, 89081 Ulm, Germany}

\author{Toma Susi}
\affiliation{University of Vienna, Faculty of Physics, Boltzmanngasse 5, 1090 Vienna, Austria}

\author{Maxime Debiossac}
\affiliation{German Aerospace Center (DLR), Institute of Quantum Technologies, Wilhelm-Runge-Stra\ss e 10, 89081 Ulm, Germany}

\author{Philippe Roncin}
\affiliation{Institut des Sciences Mol\'eculaires d'Orsay (ISMO), Centre national de la
recherche scientifique (CNRS), University Paris-Sud, Universit\'e Paris-Saclay, Orsay, France}

\author{Christian Brand}
\email{Christian.Brand@dlr.de}
\affiliation{German Aerospace Center (DLR), Institute of Quantum Technologies, Wilhelm-Runge-Stra\ss e 10, 89081 Ulm, Germany}

\date{\today}

\begin{abstract}
Analytic methods based on matter-wave diffraction are a cornerstone in condensed-matter research, providing access to static and dynamic materials properties down to the atomic level.
In these experiments, the shape of the diffraction pattern is largely determined by the lattice at equilibrium whereas vibrationally-induced distortions are treated perturbatively. 
Here, we show that the perturbative approach does not hold for helium diffracted at kiloelectronvolt energy through freestanding single-layer graphene.
In this case, we enter a new regime of strong coupling where the projectile strongly interacts with the electron density of several lattice atoms simultaneously, leading to phase shifts of several radians.
In consequence, lattice distortions introduce a significant phase spread that cannot be described by the typically employed Debye--Waller factor.
We show that the weak-coupling regime is retained for atomic hydrogen diffraction. 
The experimental results are supported by simulations, providing a regime-independent approach to describe the influence of phonons on atom diffraction phenomena.
\end{abstract}

\maketitle
\section{Introduction}

Diffraction experiments using matter waves or X-rays provide information about structures reaching down to the scale of single atoms and molecules. 
In these experiments, each diffracted species exhibits a different interaction mechanism with the materials, allowing the probe to be tailored to the respective scientific question.
Whereas X-rays~\cite{warren1990x}, electrons~\cite{cowley1992electron}, positrons~\cite{Rosenberg_JVacSciTechnol17_253, hugenschmidt2016positrons}, and atoms~\cite{farias1998atomic, Bracco-Holst_SurfaceScienceTechniques_2013} are sensitive to the electron distribution in the material, neutrons interact only with the atomic nuclei~\cite{bacon1975neutron}.
Also the depth probed in the material varies among these techniques: 
neutrons and X-rays are typically used to probe bulk materials while electron and positron diffraction excel in the study of thin materials.
Finally, atoms and molecules interact only with the topmost layer of the surface~\cite{Estermann_ZPhys61_95, Bracco-Holst_SurfaceScienceTechniques_2013, Winter_ProgSurfSci86_169, Cronin_RevModPhys81_1051}. 
The large diversity of particles is directly connected to a wide range of applications in structural biology~\cite{Yip_Nature587_157, GonencryoTEM_2022, Chapman_Nature470_73, Tenboer_Science346_1242, deJonge_NatNanotechnol6_695}, chemistry~\cite{Liu_Nature551_494, Zhou_Nature621_75, Alcorn_NatRevChem7_256}, and materials research~\cite{Plotkowski_NatCommun14_4950, Chi_PhysRevLett117_047003, Farias_RepProgPhys61_1575, Bracco-Holst_SurfaceScienceTechniques_2013, holst2021material}.
Each technique has a range of experimental parameters where it works best, and combining these different information channels reveals highly detailed insights into the structure and properties of materials.

What is common to all these techniques is that the interaction between the wave and an individual scattering center is weak and fast.
This is due to weak interaction potentials, high particle velocities, or a combination of both.
In consequence, the impact of vibrationally-induced lattice distortions on the diffraction pattern can be treated as a small perturbation and is typically well-described via the Debye--Waller factor (DWF)~\cite{manson1991inelastic, Manson_SurfSciRep2022}.

In this study, we show that by diffracting helium with energies around 1~keV through single-layer graphene, we leave the realm of perturbative (weak) coupling and enter a new regime of strong coupling. 
In this regime, the diffracted particle acquires substantial phase shifts from more than one lattice atom simultaneously.
In consequence, lattice distortions have a significant impact on the particle's overall phase and lead to a position-dependent phase spread.
This results in a drastic rearrangement of the relative peak intensities in the diffraction pattern including an apparent violation of the material's structure factor.

Thus, although the material is only one atomic layer thin, we are not within the thin-grating approximation where the diffraction pattern is connected to the equilibrium structure of the lattice via a Fourier transform.
In consequence, also the standard theoretical description using the Debye--Waller factor fails to reproduce the essential features of the diffraction pattern.

However, we can accurately describe the experimental findings by simulating the diffraction pattern based on a large phonon-distorted lattice. 
Moreover, we find that the phase spread significantly reduces the area of the lattice compatible with coherent diffraction.
For hydrogen atoms in the same energy range, the acquired phase shift is smaller by about an order of magnitude and we retain diffraction in the weak-coupling regime.

\begin{figure*}[t]
    \centering    \includegraphics[width=0.98\textwidth]{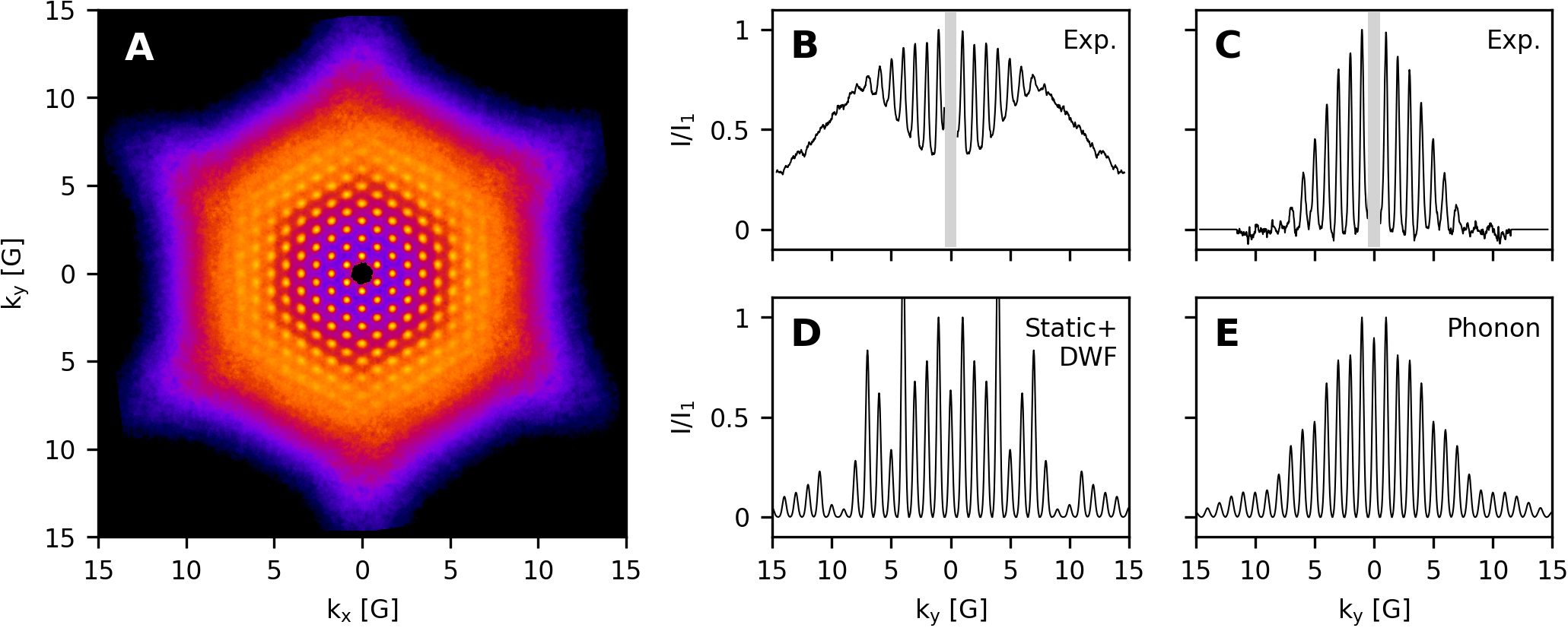}
    \caption{\textbf{Diffraction of He atoms in the strong-coupling regime}. 
    (\textbf{A}) Experimental diffraction pattern (symmetrized with 60\textdegree\ rotation) of He atoms at 907~eV kinetic energy through suspended monocrystalline monolayer graphene. 
    The diffraction pattern is plotted in units of transferred grating momenta $G=4\pi/(\sqrt{3}a)$, with $a=246$~pm, the lattice constant of graphene. 
    The direct beam is blocked in the center of the pattern. 
    Central vertical full profile (\textbf{B}) and profile with the inelastic background subtracted (\textbf{C}) 
    of the diffraction pattern in (A). 
    (\textbf{D}) Simulation of the diffraction profiles based on a lattice at equilibrium with the Debye--Waller factor to account for thermal displacements. 
    (\textbf{E}) Background-subtracted simulation of the elastic diffraction profiles including the phonon-induced distortions. 
    In (B) and (C) the grey-shaded area marks the position of the beam block in the experiment. 
    In all panels the intensity has been normalized to the first diffraction order.
    }
    \label{fig:fig_He}
\end{figure*}

\section{Diffraction through monocrystalline graphene}

We diffract helium and hydrogen atoms at kiloelectronvolt energies through suspended monocrystalline single-layer graphene, see Methods and Ref.~\cite{kanitz2025diffraction} for more details on the experimental setup. 
The atom beam is produced by first creating ions, accelerating them to the desired kinetic energy $E_0$, and then neutralizing them by near-resonant charge transfer~\cite{Rousseau_PhysRevLett98_016104}. 
The beam is collimated before being diffracted through the graphene sample and the pattern is recorded on a position-sensitive detector. 
The diffraction angle $\theta$ is given by the ratio of imparted grating momentum $\hbar G$ and forward momentum $\hbar k=h/\lambda_{\mathrm{dB}}$ with the de Broglie wavelength $\lambda_{\mathrm{dB}}=h/\sqrt{2mE_0}$.
Here, $h$ is Planck's constant and $m$ the mass of the atom.

The diffraction pattern of He at a kinetic energy of $E_0=907$~eV is shown in Fig.~\ref{fig:fig_He}A.
Contrary to diffraction through polycrystalline graphene that exhibits Debye--Scherrer rings~\cite{kanitz2025diffraction}, the pattern consists of well-resolved peaks arranged in a trigonal pattern.
Those peaks are sitting on top of a broad star-shaped signal, which mirrors the symmetry of the grating, see Fig.~\ref{fig:fig_He}A and B.
Previously, it was shown that diffraction of H through graphene is elastic with a broad signal component associated with electronically inelastic scattering, that is, energy loss of a few electronvolts~\cite{Guichard_PhysRevLett135_263403}. 
As such a coupling is also predicted for helium~\cite{Brand_NewJPhys21_033004, kanitz2025diffraction}, we associate the diffraction peaks with elastic contributions and ascribe the broad signal to atoms that underwent inelastic scattering. 
This is denoted as background in the following. 

After background subtraction (see Methods), we obtain the experimental elastic peak intensities. 
The vertical ($y$) cut through the center of the diffraction pattern is shown in Fig.~\ref{fig:fig_He}C, for the horizontal ($x$) cut refer to the Supplementary Information.
The pattern exhibits well-resolved peaks up to the seventh diffraction order whose intensities seem to be governed by a Gaussian envelope. 
This observation is in stark contrast to the theoretical predictions based on the lattice near equilibrium (see Fig.~\ref{fig:fig_He}D), where interference within the lattice cell strongly modulates the diffracted intensity.

The diffraction pattern shown in Fig.~\ref{fig:fig_He}D was calculated within the eikonal approximation, applying the Debye--Waller factor (DWF) to account for the effect of phonon-induced lattice distortions. 
The DWF describes the reduction of the diffracted intensities by the factor $\exp{(-\langle{\delta \phi}^2\rangle)}$, where $\delta\phi = \Delta \mathbf{k}\cdot\mathbf{u}$ is the phase shift associated with the lattice atoms' thermal displacement $\mathbf{u}$ and the momentum transfer $\Delta\mathbf{k}$ of the impinging atom.
This approach is widely employed to describe the effect of vibrations on the diffraction patterns of X-ray, atoms, electrons and neutrons.

\section{Strong-coupling regime}

\begin{figure*}
    \centering
    \includegraphics[width=0.65\linewidth]{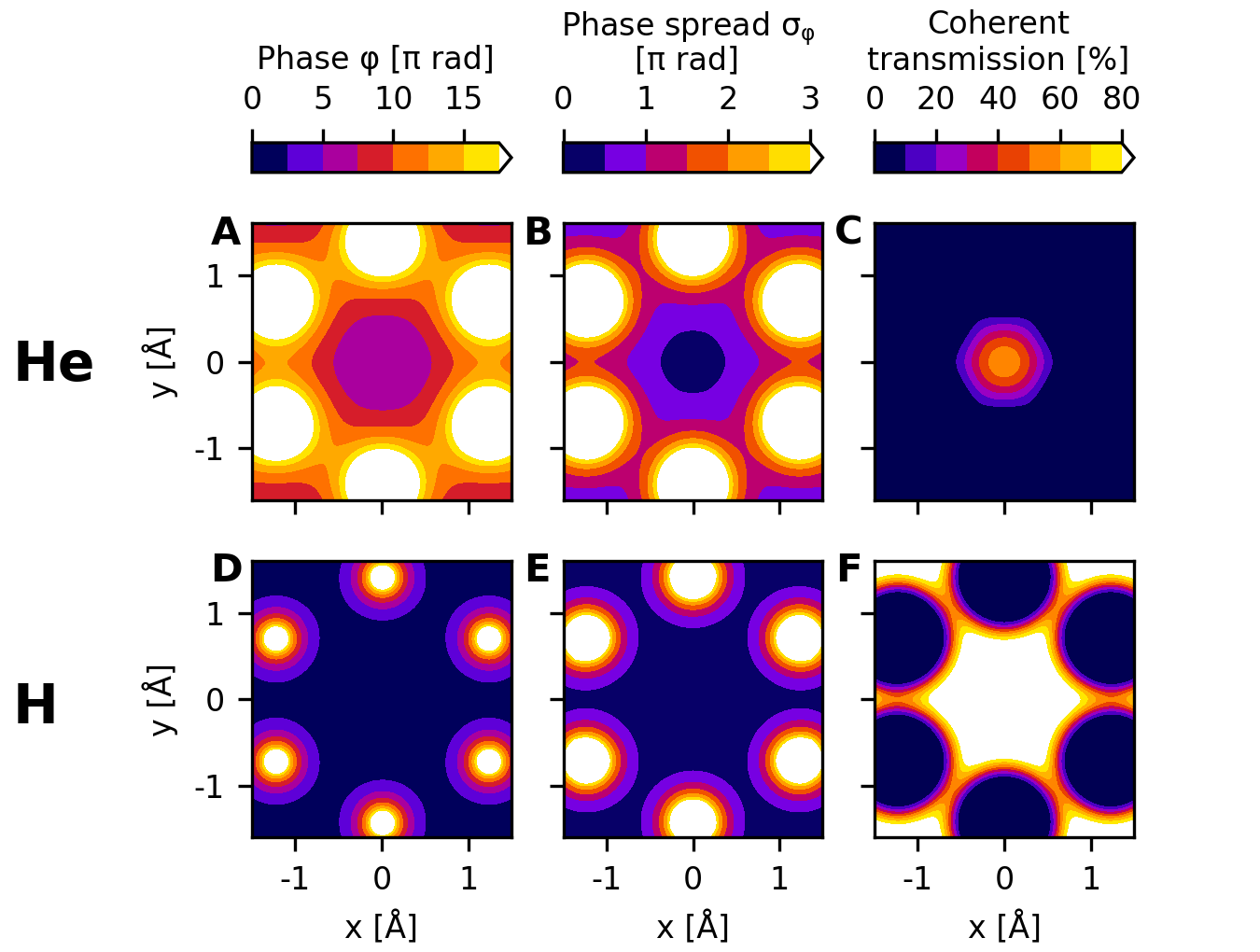}
    \caption{\textbf{Phase landscapes of the 
    lattice in the strong- and weak-coupling regime}. (\textbf{A} and \textbf{D}) Phase for helium at 907~eV (A) and hydrogen at 928~eV (D) passing through single-layer graphene. (\textbf{B} and \textbf{E}) Standard deviation of the phase shift experienced by helium (B) and hydrogen (E) due to coupling to the lattice vibrational modes. (\textbf{C} and  \textbf{F}) Coherent transmission factor for helium (C)  and hydrogen (F) given by $\exp(-\sigma_\phi^2/2)$, see Supplementary Information.}
    \label{fig:phase_landscape}
\end{figure*}

To understand the disagreement between experiment and theory, we examine the area of validity of the applied models, that is, the eikonal approximation and the Debye--Waller factor. 
The eikonal approximation is well justified considering the high kinetic energy $E_0$ of the projectile at 907~eV (see Supplementary information). 
This is much larger than the interaction potential $V(x,y,z)$ anywhere but directly adjacent to the grating atoms, and $\lambda_\mathrm{dB}$ (480~fm) is much smaller than the interaction range~\cite{barone_2002_high_energy}.

The Debye--Waller factor, on the other hand, relies on two assumptions~\cite{levi1979quantum}: the interaction is (i) weak and (ii) fast. 
While the latter is valid for light atoms at keV-energy, the former implies that changing the lattice atoms' position introduces only a small perturbation to the diffracted particle's phase.
To verify whether the assumption of weak interaction is justified in the present situation, we first calculate the accumulated phase based on He--graphene potentials derived from density functional theory (DFT) simulations (see Supplementary Information).
The potential landscape shown in Fig.~\ref{fig:phase_landscape}A illustrates that the phase for He at 907~eV never drops below $5 \pi$ rad, even in the hexagon's center.
Thus, all six surrounding carbon atoms have a significant contribution to the overall phase $\phi$. Notably, these phases are two to three orders of magnitude larger than for electrons at energies typical for transmission electron microscopy (Supplementary Information).
The high phase for He--graphene stands in contrast to the situation of H--graphene at a comparable energy (Fig.~\ref{fig:phase_landscape}D), where significant phase is only acquired close to a C atom.

In a second step, we investigate temperature-induced distortions of the hexagons to assess their impact on the phase.
For each distorted lattice geometry, we calculate the position-dependent value of $\phi$.
Sampling over a large number of geometries thus allows us to extract the position-dependent phase spread $\sigma_\phi$ in the hexagon.
Again, we observe a strong difference between H and He, see Fig.~\ref{fig:phase_landscape}B and E.
For hydrogen, $\sigma_\phi$ is less than $\pi/2$ over large parts of the hexagon, 
suggesting that phonons have little influence on the phase. 
However, for helium only a small
region in the hexagon's center is weakly sensitive to the lattice distortions. 
For instance, the atomic trajectory passing through the C--C bond position experiences so much dephasing that it effectively no longer interferes with the trajectory passing through the hexagon center (Supplementary Information).
This translates into an area of coherent transmission, which is strongly reduced for He compared to H, as shown in Fig.~\ref{fig:phase_landscape}C and F.

Thus, while requirement (i) of the DWF is fulfilled for H, it is violated for He at the considered energy.
In consequence, the impact of the phonons on the diffraction pattern of helium cannot be described perturbatively but has to be modeled explicitly. 

\section{Impact of phonons on the pattern}

To model diffraction in the strong-coupling regime, we start from a large crystal covering $96\times 96$ hexagons and introduce thermal distortions at $T=300$~K, which are dominated by the zero-point motion (see Methods). 
Attaching the fitted binary He--C potential $V(x,y,z)$ to each lattice atom provides us with the phase mask of the distorted lattice that is used to calculate the diffraction pattern within the eikonal approximation.
To accumulate sufficient statistics, we averaged the results from 100 distorted lattice snapshots.

The results in Fig.~\ref{fig:fig_He}E show a much better agreement with experimental data (Fig.~\ref{fig:fig_He}C) than the prediction based on a lattice at equilibrium modulated by the DWF. 
As in the experiments, we observe a monotonous decrease in relative intensity with increasing momentum transfer. 
Moreover, the width of the envelope is well reproduced given that the simulation has no free parameters.
We note that the simulations predict some remaining intensities for ($k_{x,y}\geq10~G$), which is not present in the experiment. 
This is probably due to electronic decoherence that is not included in the simulations.

\begin{figure*}[t]
\centering    \includegraphics[width=0.95\textwidth]{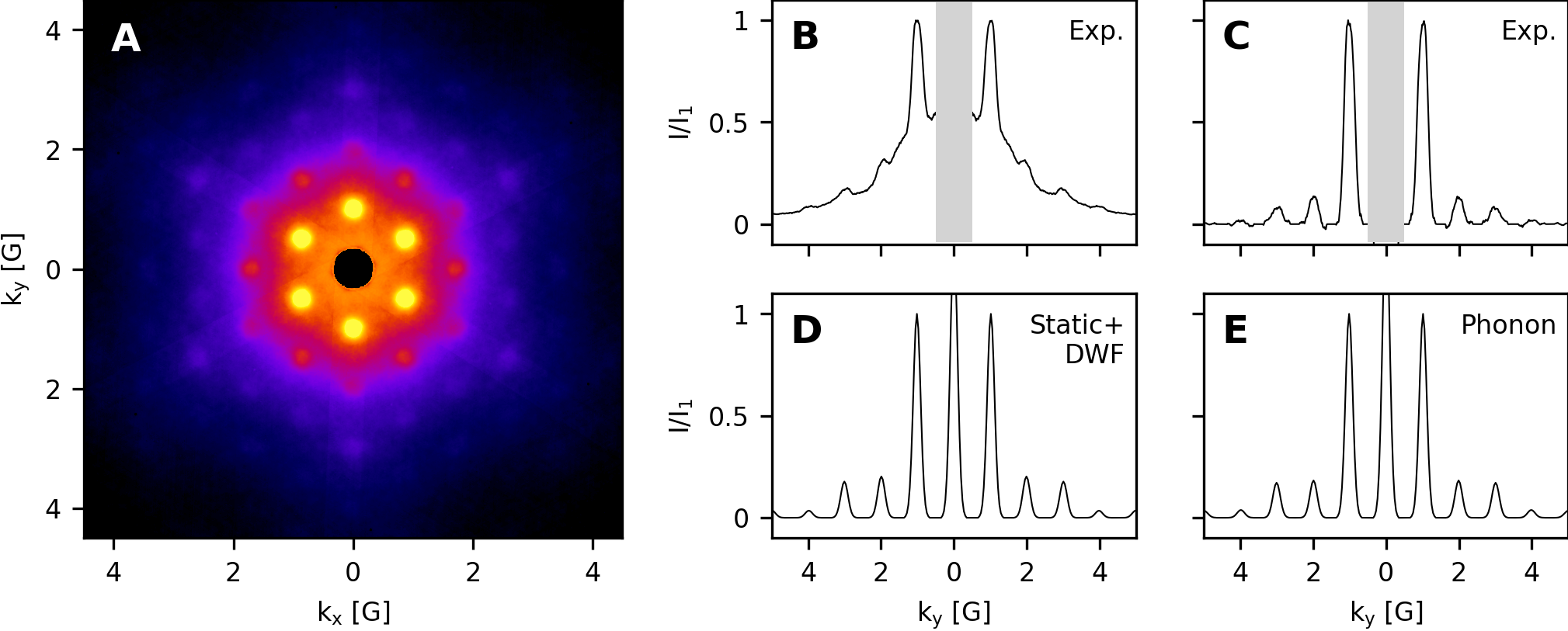}
\caption{\textbf{Diffraction of H atoms in the weak-coupling regime}. (\textbf{A}) Experimental diffraction pattern (symmetrized with 60\textdegree\ rotation) of neutral H atoms at 928~eV kinetic energy through suspended monocrystalline monolayer graphene. 
The direct beam is blocked in the center of the pattern. 
(\textbf{B} and \textbf{C}) Vertical profiles through the center of the diffraction pattern in (A) with and without subtraction of the inelastic background. 
(\textbf{D}) Corresponding simulation of the elastic diffraction profiles using Debye--Waller theory. 
(\textbf{E}) Simulation of the elastic diffraction profiles including the phonon modes. 
In panels B and C the grey-shaded area mark the position of the beam block in the experiment. 
In all panels the intensity has been normalized to the first diffraction order.}
    \label{fig:fig_H}
\end{figure*}

The simulations also explain the apparent deviation from the material's structure factor, which describes how the lattice geometry influences the relative intensities of diffraction orders in a transmission experiment~\cite{warren1990x}.
In graphene, this is determined by the two trigonal sub-lattices, which modulate the intensities in the pattern by a factor of four~\cite{shevitski2013dark}. 
This is not observed in the experiments, see Fig.~\ref{fig:fig_He} and Supplementary Information.
The structure factor assumes diffraction from multiple independent scattering sites inside the unit cell. 
However, due to the strong phase fluctuations induced by the distortions, the scattering sites relevant for diffraction are now the hexagon centers.  
In consequence, instead of diffracting through a honeycomb lattice defined by atoms, the matter wave is effectively diffracted through a trigonal lattice of hexagon centers.
For such a grating geometry the structure factor is independent of the diffraction order, which is in agreement with the experiment.
Thus, the structure factor is not violated, but instead the geometry of the scattering element has changed, see Supplementary Information.

\section{Application to the weak-coupling regime}

To test the general validity of the theoretical approach, we apply it to diffraction of atomic hydrogen at a comparable energy.
As diffraction of H fulfills the requirements of the DWF, we expect it to be in the weak-coupling regime.
Similarly to He, we have calculated the diffraction pattern within the eikonal approximation based on H--graphene DFT potentials. 

Figure~\ref{fig:fig_H} shows the comparison between the experimental results, the prediction based on the lattice at equilibrium with the DWF, and the explicit phonon simulation.
The diffraction pattern only extends to a few reciprocal vectors $G$ as the interaction is not strong enough to populate higher diffraction orders.  
Comparing the experimentally observed intensities (panel C) with the simulations shows excellent agreement with both levels of theory.
This confirms that diffraction of hydrogen through graphene is in the perturbative-interaction regime and that the simulations based on the distorted lattice are equally well suited to describe this situation.

\begin{figure*}
    \centering
    \includegraphics[width=\linewidth]{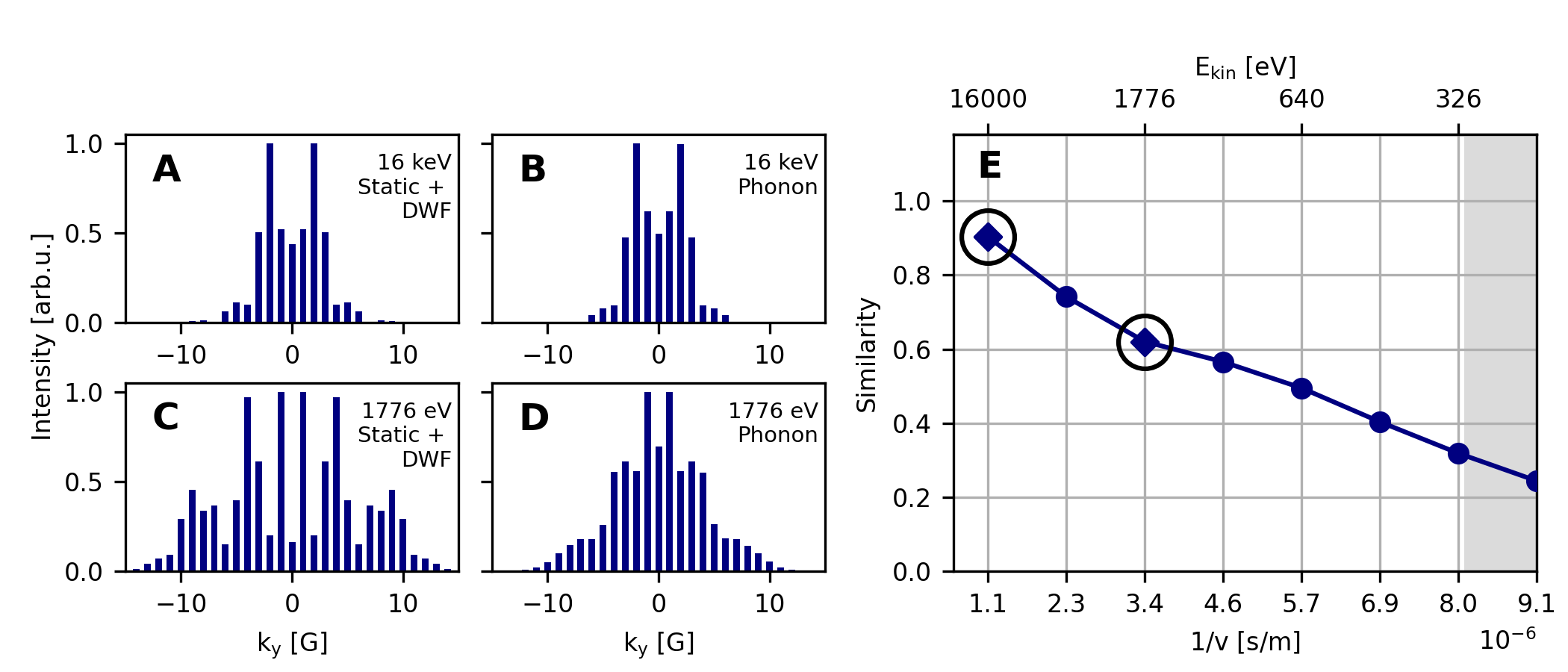}
    \caption{\textbf{Simulations of diffraction peak intensities at varying energies using two different methods}. (\textbf{A} and \textbf{C}) Peak intensities for helium diffraction at 16\,keV  and 1776\,eV generated from the static lattice geometry and modulated by a DWF. (\textbf{B} and \textbf{D}) Helium simulations at 16\,keV and 1776\,eV averaging over the diffraction images of 100 phonon-distorted lattice configurations.
    (\textbf{G}) Similarity score $S$ of the phonon and DW simulations at varying kinetic energies and plotted as a function of $1/v$ to highlight the effect of the interaction time.
    The circled data points designate the data shown in panels (A-D). 
    The grey area designates the breakdown of the eikonal approximation (Supplementary Information).}
    \label{fig:simulation_compare}
\end{figure*}

To estimate at which point the explicit-phonon model converges to the Debye--Waller model, we 
simulated the diffraction of helium and hydrogen over a large energy range using both methods.
Figure~\ref{fig:simulation_compare}A-D shows exemplary diffraction patterns generated for helium using both simulation methods at two different energies. 
At 1776\,eV the patterns have few similarities.
This is mirrored by the similarity score $S$ in Fig~\ref{fig:simulation_compare}, which compares the relative intensities of the elastic peaks for both simulations (see Methods). 
With increasing energy of the particle, the images become progressively more similar.
At 16~keV $S$ is around 90\% for He and it can be clearly seen that both methods yield similar predictions for the relative peak intensities. 
To highlight the effect of the interaction time, we also plot $S$ depending on $1/v$.

For hydrogen, the similarity score lies above 90\% over the whole considered energy range, see Supplementary Information. 
This suggests that at these energies H remains in the perturbative coupling regime.

\section{Discussion and Outlook}

We observe for the first time atomic diffraction in the strong-coupling regime.
In this interaction regime, only some parts of the unit cell contribute to the diffraction pattern, while the influence of other parts is heavily suppressed due to large phase fluctuations emerging from the thermal motion of graphene. 
This leads to a dramatic rearrangement of intensity within the diffraction pattern.
In consequence, the Debye--Waller theory that is very successful in reproducing experimental diffraction patterns since a century~\cite{Farias_RepProgPhys61_1575, Manson_SurfSciRep2022, warren1990x, rousseau_JournalOfPhysics2008_012013} does not hold in the strong-coupling regime.
To reproduce the experimental results, we explicitly took phonon-induced lattice distortions into account.
As demonstrated for H and He, this parameter-free theoretical model correctly describes diffraction both for weak and strong coupling.
The remaining deviations between theory and experiment for He--graphene point towards the influence of electronically inelastic interactions~\cite{Jiang_Science364_379, Bunermann_JPhysChemA125_3059, Guichard_PhysRevLett135_263403}.
The future study of such couplings thus promises insights on position-dependent interactions between the projectile and the electronic structure of materials. 

The strong-coupling regime exists between two extremes. 
On the one side, there is classical scattering, where diffraction is completely suppressed due to high phase spread everywhere in the unit cell~\cite{Frank_Carbon257_121746_2026, Buhler_FrontChem11_1291065}. 
On the other side, there is transmission with only low phase fluctuations where the entire unit cell contributes to the diffraction pattern. 
In this case, only slight modifications such as the DWF need to be used to include the effect of thermal displacements afterwards.
Thus, we are exploring a quantum-to-classical transition, where the matter wave is dephasing due to the strength of the interaction of the particle with the grating structure. 

\section*{Acknowledgement}

This study was supported by Austrian Science Fund grant P 36264-N (V.Z. and T.S.). 
Part of the computational results have been achieved using the Austrian Scientific Computing (ASC) infrastructure.

\bibliographystyle{apsrev4-2}
\bibliography{References}

\newpage

\section{Methods}

\subsection{Experimental setup}

A more detailed description of the general experimental setup can be found in Ref.~\cite{kanitz2025diffraction}. 
Beams of singly-charged helium or hydrogen atoms are created from pure helium or hydrogen gas in a commercial electron-impact ion gun, including collimation and steering optics.
For the creation of hydrogen ions, a Wien filter is used to ensure the beam consists only of atomic ions. 
The created ion beam is neutralized in a gas-filled cell via (near-)resonant charge exchange on argon for hydrogen and helium for helium at a pressure of $6\cdot10^{-2}\,\text{mbar}$. 
The ion-beam intensity can be monitored by a retractable Faraday cup downstream of the charge-exchange cell. 
The beam is collimated by two pinholes with variable diameters that enclose the charge-exchange cell, as well as by a third pinhole with variable diameter located 128~mm in front of the sample.

The graphene sample consists of a monolayer of monocrystalline graphene grown by chemical vapour deposition. 
It is suspended on a holey carbon film with 1.2~\textmu m sized holes arranged with a center-to-center spacing of 2.5~\textmu m. Both layers are on top of a gold TEM grid resulting in a total thickness of 25 \textmu m. To roughly clean the samples, they were thermally annealed by heating them to 450\textdegree C for 90 minutes. Prior to each diffraction experiment, laser cleaning~\cite{Tripathi_pssr11_1700124} was performed with a 300 mW laser beam at 532~nm, exposing the sample two times for 20 seconds with 60 seconds cooling time in-between.

After interacting with the graphene sample, the atom beam travels freely for 709~mm before impinging onto a dual-plate multi-channel detector (MCP) stacked onto a phosphor screen. The image on the phosphor screen is recorded by a CMOS camera and the direct beam is dumped using a beam block in front of the MCP. 

\subsection{Image correction and post-processing}
Images are recorded and processed analogously to the method described in Ref.~\cite{kanitz2025diffraction}. Additionally, images are corrected for inhomogeneities in the detector sensitivity. 
To determine the correction pattern, events from remaining background gas in the vacuum chamber were recorded over multiple nights at a high MCP gain.

The diffraction pattern position, scale, and orientation was determined by fitting a 2D pattern of Gaussian peaks in a trigonal lattice arrangement to the middle part of the pattern.
Using the position value from the fit, the effective shot-noise of the images was improved by taking advantage of the six-fold symmetry of the lattice and averaging analogous parts of the pattern.

In order to separate the diffraction peaks from the diffuse inelastic contributions, the peak areas were removed from one copy of the image based on the fit results. 
The blank areas were then filled by linear interpolation. 
The background image created in this way was then subtracted from the full image to generate the bare peaks from which slices were taken to generate the data for Figs. \ref{fig:fig_He}C and \ref{fig:fig_H}C.

\subsection{Simulation of diffraction patterns}

At the energies considered in this work, the projectile–graphene interaction is so brief that the carbon atoms do not have time to move and can therefore be treated as fixed. 
Within this so-called sudden approximation (also known as the frozen-phonon approximation in TEM), the calculation of projectile diffraction reduces to the calculation of projectile scattering by the potential generated by the fixed carbon nuclei.
This is done using the eikonal approximation. 
For a discussion of the validity of this approximation in the current case, see Supplementary Information.

The angular scattered intensity calculated from the eikonal phase $\phi_\text{eik}$ can be expressed as
    \begin{align}
        I(\bm \kappa) &=
        \frac{1}{A} 
        \left|
        \iint_{\bm A}
        \exp ({i\phi_\text{eik}(\bm s)}) ~
        \exp(-i \bm \kappa \bm s) ~
        \text{d}^2 \bm s ~
        \right|^2
\label{Eq:FT}\\
        \phi_\text{eik}(\bm s)&=
        -\frac{1}{\hbar v_0}
        \int V(\bm s,z) ~ \text{d}z,
    \end{align}
where $z$ is the direction along the beamline, $\bm s = (x,y)^T$ is the impact parameter, $\bm \kappa = (k_x,k_y)^T$ is the momentum transferred in the in-plane direction, $\bm A$ is the graphene surface, $A$ its area, and $v_0$ is the atom velocity. 

The eikonal approximation only holds when the transferred momentum $\left | \bm \kappa \right|$ is much lower than $k$, i.e. at small values of $\theta$.

The Fourier transform is carried out using $128 \times 128$ impact parameters per unit cell.
When considering a lattice at equilibrium, it is sufficient to integrate over one periodic unit cell. 
In that case, the discrete Fourier transform of Eq.~\eqref{Eq:FT} gives the intensity of $128 \times 128$ elastic peaks.
When the lattice is deformed, we integrate over a large periodic supercell, built with $96 \times 96$ unit cells, thus involving $(96 \times 128) \times (96 \times 128)$ impact parameters. Intensity then appears at additional positions in the diffraction image corresponding to inelastic processes. 
These inelastic peaks are $96 \times 96$ times more numerous than the elastic peaks.

To account for sources of experimental broadening, each of the discrete peaks of zero width characterized by $\kappa$ is replaced by a circular Gaussian spot whose width $\delta \kappa$ comes from three contributions:
\begin{equation}
\delta \kappa = \sqrt{ 
\delta \kappa_T^2 + \delta \kappa_L^2 + \delta \kappa_0^2}.
\end{equation}
The limited transverse coherence length of the atomic beam $\ell_T$, as generated from a $d=500$~\textmu m aperture at $L=900$~mm upstream of the lattice, is
\begin{equation}
\ell_T = \frac{4 \pi L}{kd}
\end{equation}
that induces a broadening
\begin{equation}
\delta \kappa_T = \frac{2 \pi}{\ell_T}  = \frac{kd}{2L}~.
\end{equation}
The longitudinal coherence of the atomic beam, related to the energy spread $\delta E = 20$~eV based on previous source characterizations, induces a broadening
\begin{equation}
\delta \kappa_L = \frac{\delta E}{2E} ~ \kappa~.
\end{equation}
Finally, the finite collimation angle of the atomic beam $\delta \theta_0$ observed in the experiment ($\delta \theta_0 = 0.7$~mrad for helium, and 1~mrad for hydrogen) induces an angular broadening
\begin{equation}
\delta \kappa_0 = k \delta \theta_0~.
\end{equation}
Note that at a fixed energy $E_0$, the broadening related to the transverse coherence and to the beam aperture does not depend on the peak position, in contrast with the broadening due to the energy spread of the beam, which increases at large $\left| \bm \kappa \right|$.

\subsection{Phonon-displacement generation}

The planar graphene specimen model consists of $n_\mathrm{cell}$ = 96$\times$96 elementary cells, each containing two carbon atoms. 

When the projectile is transmitted perpendicularly through the sheet, the contribution of the carbon atoms' out-of-plane movement to the phase shift can be neglected. 
Thus, we only consider the in-plane motion of the C atoms.
The vector $\mathbf{R}$, representing the set of their positions, then has $n = 4n_\mathrm{cell} = 36864$ free coordinates.
We determine the equilibrium position $\mathbf{R}_0$ of the C atoms in the graphene membrane by solving the $n\times n$ Hessian matrix
\begin{align}
    \mathcal{H} = 
    \left[ {\begin{array}{cccc}
        \frac{\partial^2 V}{\partial q_1^2} & \frac{\partial^2 V}{\partial q_1 \partial q_2} & \hdots & \frac{\partial^2 V}{\partial q_1 \partial q_n} \\
        \frac{\partial^2 V}{\partial q_2 \partial q_1} & \frac{\partial^2 V}{\partial q_2^2} & \hdots & \frac{\partial^2 V}{\partial q_2 \partial q_n} \\
        \vdots & \vdots & \ddots & \vdots \\
        \frac{\partial^2 V}{\partial q_n \partial q_1} & \frac{\partial^2 V}{\partial q_n \partial q_2} & \hdots & \frac{\partial^2 V}{\partial q_n^2}
    \end{array}} \right]
\end{align}
based on the C--C interaction potential $V(\mathbf{R})$ that is calculated using a REBO (Reactive Empirical Bond Order) method~\cite{Stuart_ChemPhys112_6472}, 
Here, $q_1, q_2,...,q_n$ are the coordinates of the C atoms. 
The eigenvalues of $\mathcal{H}$ are denoted as $m_C\omega_k^2$, where $m_C$ is the mass of the carbon atom. 
Its normalized eigenvectors, denoted $\mathbf{e}_k$, are the vibrational modes of the sheet. 

The first two lowest-frequency solutions describe global translations of the sheet in its own plane, and are disregarded. 
The remaining $n-2$ modes correspond to the graphene lattice distortion. 
To build the distorted lattice, we first randomly fix the quantum vibrational numbers $v_k$ associated with each mode $k$, such that the sampling respects Bose--Einstein statistics, that is, its probability is given by
\begin{equation}
    p(v_k) = \left[\exp\left(\left(v_k+\frac{1}{2}\right)\frac{\hbar\omega_k}{k_BT}\right)-1\right]^{-1}.
\end{equation}
We then fix the displacement of the C atoms due to each $k$ mode by
\begin{equation}
    \delta \mathbf{R}_k = x_k\mathbf{e}_k,
\end{equation}
where $x_k$ is randomly chosen to reflect the probability density
\begin{equation}
    |\psi_{v_k}(x,\omega_k)|^2 = \sqrt{\frac{1}{\pi\sigma^2}}\left(\frac{1}{2^vv!}\right)H^2_v\left(\frac{1}{\sigma}\right)\exp\left(-\frac{x^2}{\sigma^2}\right).
\end{equation}
Here $\sigma = \sqrt{\hbar/m\omega}$ and $H_v$ is the $v^{th}$ order Hermite polynomial. Finally, the position of the C atoms is given by
\begin{equation}
    \mathbf{R} = \mathbf{R}_0 + \sum_k\delta\mathbf{R}_k.
\end{equation}

\subsection{Simulation to estimate coherence loss}

 In order to generate the phase spread in Fig.~\ref{fig:phase_landscape}, phase images were generated for 500 lattice geometries using the same binary potentials as for the diffraction simulations. The phase spread is calculated as the standard deviation of the phase over these 500 samples. For each configuration, the atoms were displaced from their equilibrium positions as independent harmonic oscillators with a mean-squared in-plane displacement of $\langle x^2\rangle=37\,\text{pm}^2$. 
 This is split into a temperature-dependent term that varies as $T\cdot\ln{(T)}$ and a zero-point motion contribution $\langle x^2\rangle_\text{zpm}=h/( 2 \pi k_Dm_Cv_s)=16$ pm$^2$ where $m_C$ is the mass of the carbon atom, $v_s=2.2\times 10^4$ m/s is graphene's in-plane speed of sound and $k_D$ is the Debye wave vector~\cite{shevitski2013dark}. 

\subsection{Similarity between diffraction simulations}

For Fig.~\ref{fig:simulation_compare}E the similarity score was calculated from the relative intensities $I_{k}$ in the elastic peaks as
\begin{equation}
    S = 1-\frac{\left<I_{ph}-I_{DW}\right>}{\left<I_{ph}\right>+\left<I_{DW}\right>},
\end{equation}
with
\begin{equation}
    \left<I\right> = \sqrt{\frac{1}{n}\sum_{k=1}^nI^2_k}.
\end{equation}
The subscript $DW$ denotes the "static+DWF" simulation and $ph$ denotes the explicit phonon simulation. 
The sum $k$ ranges over all peaks up to order 64.

\clearpage
\setcounter{figure}{0}

\renewcommand{\thefigure}{S\arabic{figure}}
\renewcommand{\thetable}{S\Roman{table}}
\renewcommand{\theequation}{S\arabic{equation}}

\section{Supplementary Information}

\subsection{Atomic potentials}

\begin{figure}[t]
    \centering
    \includegraphics[width=\linewidth]{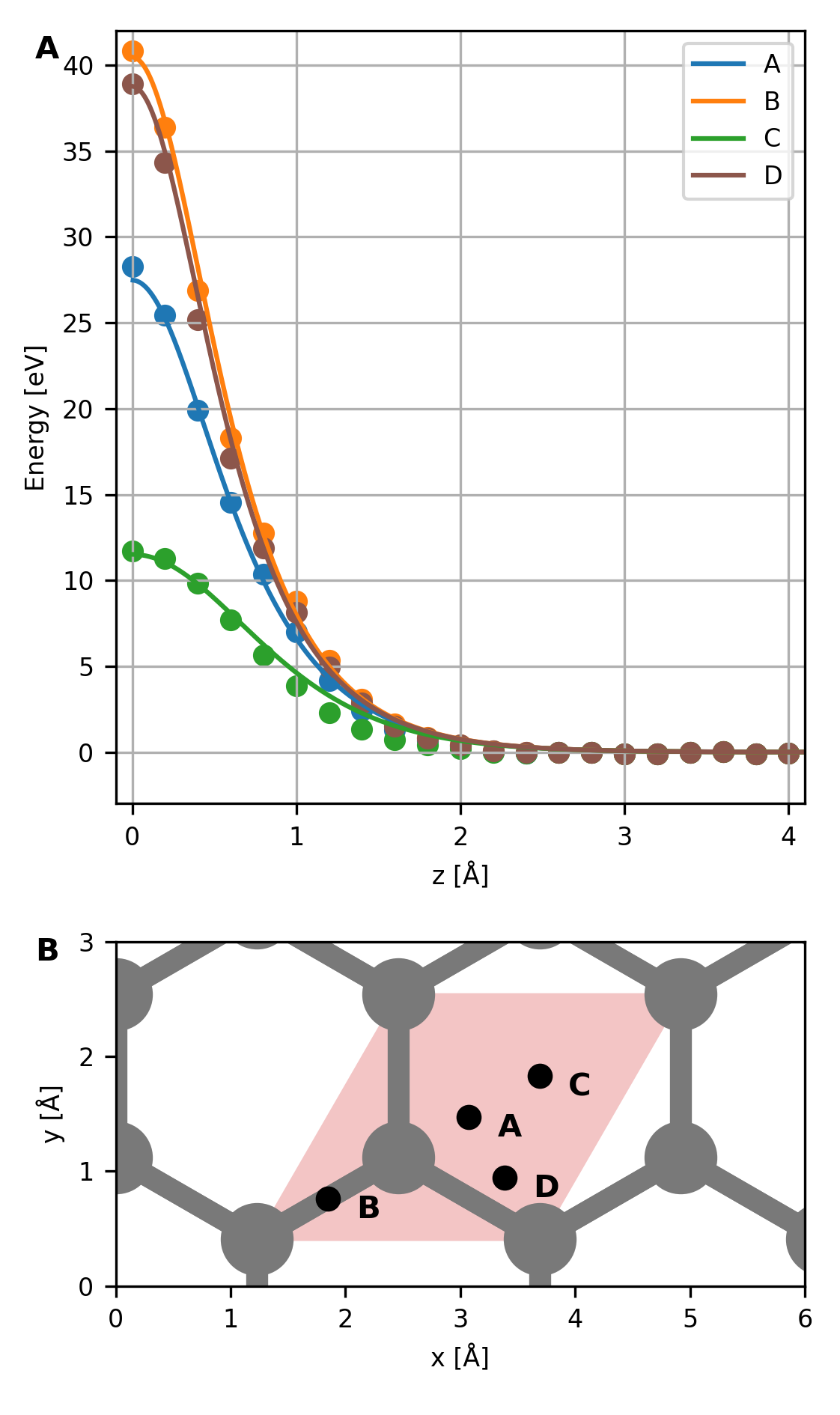}
    \caption{\textbf{A} He--graphene DFT (circles) and fitted (lines) potentials at four positions in the graphene unit cell as defined in \textbf{B}. The fitting coefficients are given in Table~\ref{tb:coefficents_fit}. \textbf{B} Definition of impact parameter points A to D. Shaded in light red is the unit cell of the lattice.}
    \label{fig:fit_DFT_He}
\end{figure}

\begin{figure}[t]
    \centering
    \includegraphics[width=\linewidth]{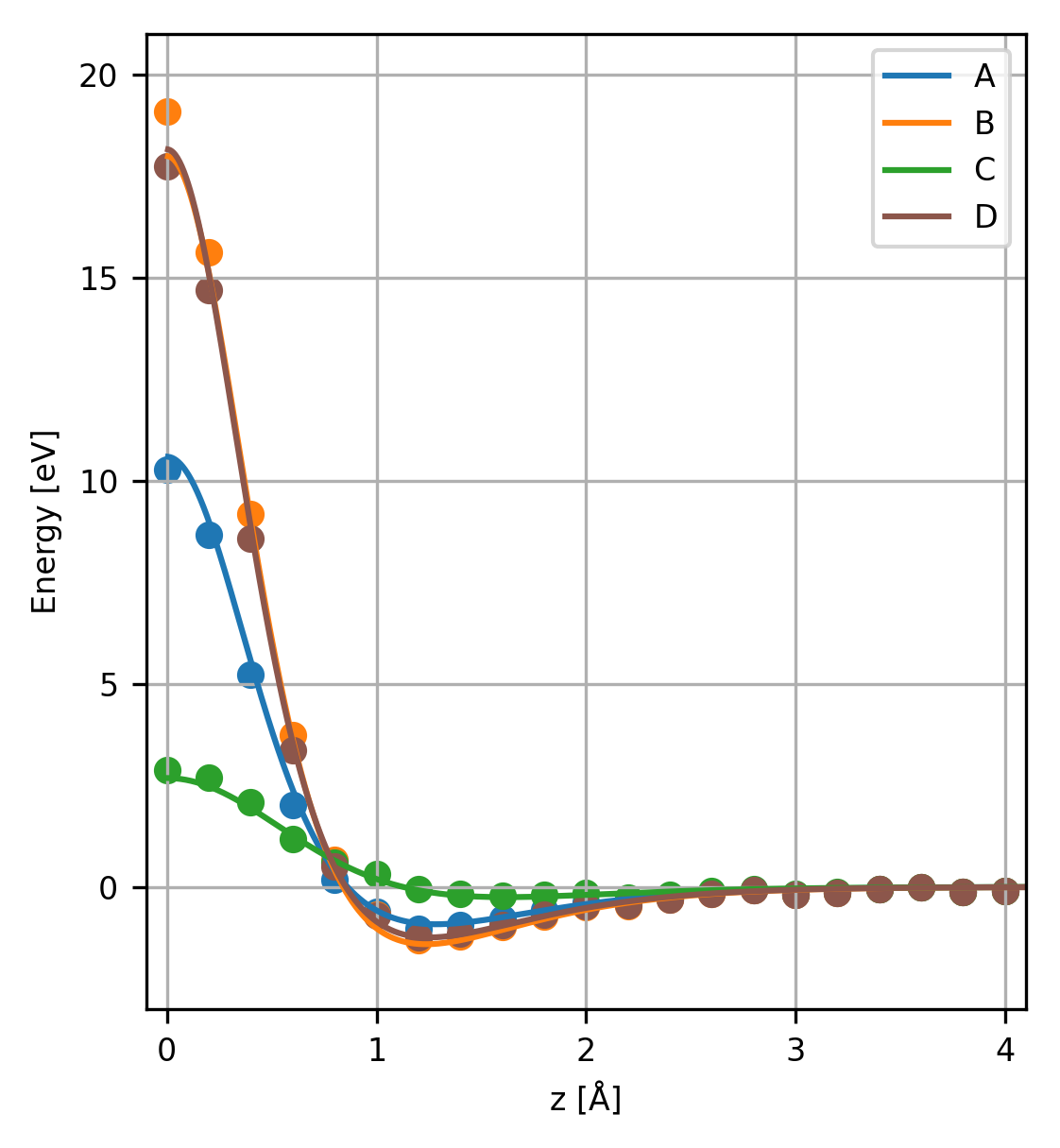}
    \caption{H--graphene DFT (circles) and fitted (lines) potentials at four positions in the graphene unit cell as defined in Fig.~\ref{fig:fit_DFT_He}B. The fitting coefficients are given in Table~\ref{tb:coefficents_fit}. The line showing the fit for position B is nearly identical to the one for position D and therefore mostly obscured.}
    \label{fig:fit_DFT_H}
\end{figure}

The static helium--graphene interaction potential $V(\mathbf{R},z)$ with $\mathbf{R}=(x,y)$ is obtained by first calculating the density functional theory (DFT) potential $V_\text{DFT}(x_n,y_n,z)$ as implemented in GPAW~\cite{Mortensen_JChemPhys160_092503, Ask_JPhysCondensMatter29_273002}.
A double-zeta polarized (dzp) basis set was employed for carbon atoms, while a single-zeta (sz) basis was used for hydrogen. 
We used a 6$\times$6 graphene supercell, and due to its sufficiently large size, Gamma-point sampling was found adequate to reach accurate results. 
All calculations were performed with the Perdew–Burke–Ernzerhof (PBE) exchange-correlation functional~\cite{perdew_generalized_1996}. 
The convergence criterion for the electron density in the initial state was set to $10^{-6}$. 
To obtain the relevant interaction potential, we calculate $V_\text{DFT}$ along a set of impact positions $(x_n,y_n)$ over the surface (Fig.~\ref{fig:fit_DFT_He}B). 
The distance between the projectile and the layer was incremented in steps of 0.2\,\AA.
For each configuration, the electronic structure was relaxed with fixed atomic positions, and the total energy of each structure was  computed and utilized to construct the potential-energy curves shown in Fig.~\ref{fig:fit_DFT_He} and Fig.~\ref{fig:fit_DFT_H}.

The discrete potential sample points are then fitted with pair-wise potentials: 
\begin{align}
    V_{\text{He--C}}(r) &= \frac{a_{1}}{r}e^{-a_{2} r} \\ 
    V_{\text{H--C}}(r,\rho) &= \frac{a_1}{r} e^{-a_2r} \nonumber\\&+e^{-\rho^2/b_1^2}\cdot b_2\left[\left(1-e^{-b_3(r-b_4))}\right)^2-1\right], \label{eq:binary_pot}
\end{align} 
where $\rho =\sqrt{(x-x_C)^2+(y-y_C)^2}$ and $r = \sqrt{\rho^2+z^2}$ are the in-plane and projectile-carbon distance respectively. 
The parameters $(x_C,y_C)$ are the in-plane positions of the carbon atoms, which are located in the $z=0$ plane. 

Whereas the helium potential can be fitted by a simple screened Coulomb model, the hydrogen potential needs to take a more complex form to accurately reproduce the attractive part of the interaction~\cite{jeloaica1999dft}.
The fit parameters $a_n, b_n$ are obtained from a least-squares fit of $V_\text{DFT}(x_n,y_n,z)$, see Table \ref{tb:coefficents_fit}.
In order to ensure fit convergence, hydrogen on trajectories close to the hexagon center needed to be given stronger weighting than those close to atoms. 
Note that the helium fit is insensitive to the weighting of the points.


\begin{table}[h!]
\caption{Parameters of the binary potential (Eq.~\ref{eq:binary_pot}) fitted on the DFT potential for He and H.}
\centering
\begin{tabular}{|c | l l l l l l|} 
 \hline
 Species &\phantom{.} $a_1$ & $a_2$ & $b_1$ & $b_2$ & $b_3$ & $b_4$ \\ [0.5ex] 

 \hline\hline
 He &\phantom{.} 68.6 & 2.29 & - & - & - & -  \\ 
 H  &\phantom{.} 66.7\phantom{.} & 1.96\phantom{.} & 1.85\phantom{.} & 25.6\phantom{.} & 2.19 \phantom{.}& 0.343 \phantom{.}\\ 
 [1ex] 
 \hline\hline
  Units &\phantom{.} eV$\cdot\angstrom$ & $\angstrom^{-1}$ & $\angstrom$ & eV & $\angstrom^{-1}$ & $\angstrom$  \\
  \hline
\end{tabular}
\label{tb:coefficents_fit}
\end{table}

The total potential $V(\textbf{R},z)$ is then reconstructed from the pair-wise potentials as $V(\textbf{R},z)=\sum_C V_{\text{H/He--C}}(r, \rho)$ where the sum runs over the positions of all carbon atoms. 
Hence, non-equilibrium lattice geometries can be taken into account by displacing the atomic centers before the full potential is constructed.
Finally, the potential is integrated along the $z$ direction to obtain $V(\textbf{R})=\int V(\textbf{R},z)dz$.

\subsection{Structure factor}

\begin{figure}[h]
    \centering
    \includegraphics[width=0.95\linewidth]{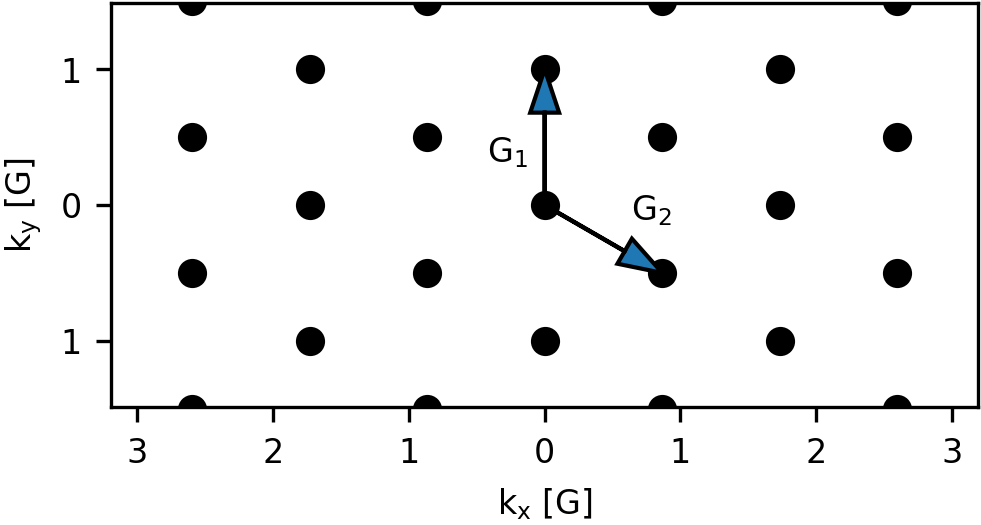}
    \caption{Trigonal reciprocal lattice unit vectors $\mathbf{G}_1$ and $\mathbf{G}_2$.}
    \label{fig:unit_vectors}
\end{figure}

Diffraction at an atomic grating can be broken down into different contributions.
The diffracting element is (if no defects are present) given by the composition of the lattice, the unit cell, and the lattice atom. In the far field and weak-coupling regime, the diffracted wave amplitude is given by the Fourier transformation of the atomic grating. 
We can use the convolution theorem to instead write the resulting amplitude as the product of the individual Fourier transformations. 
The structure factor $F$ is the diffraction contribution from the unit cell, which is given by the position of all scattering elements inside this area. 
Since we are only interested in its value at the position of diffraction orders, we use $F_{nm}$ where the integers $n,m$ designate diffraction peaks related to the momentum transfer expressed in unit lattice vectors $\mathbf{G} = n\mathbf{G}_1+ m\mathbf{G}_2$ (see Fig.~\ref{fig:unit_vectors}). 

\begin{figure}[t!]
    \centering
    \includegraphics[width=\linewidth]{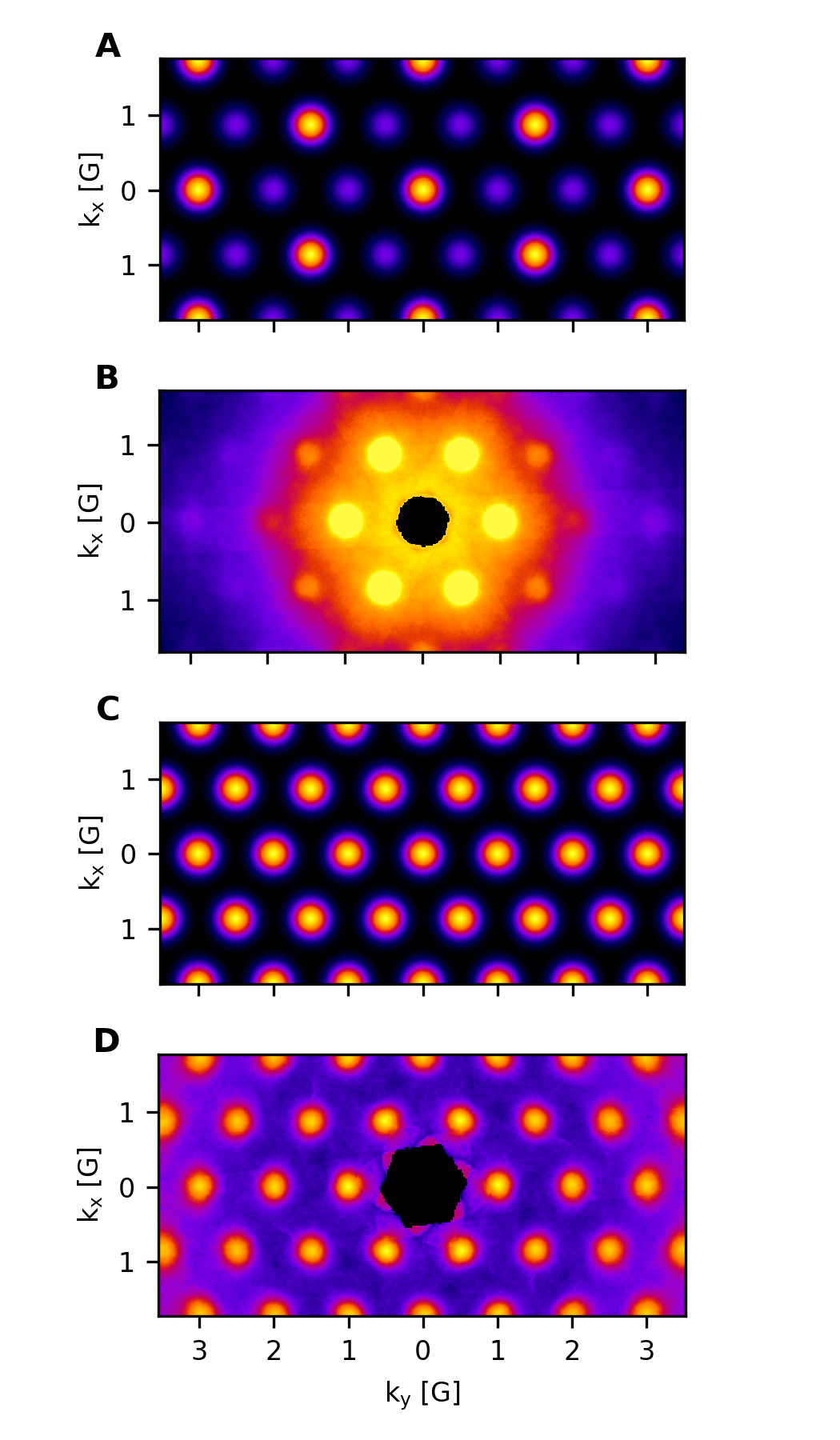}
    \caption{\textbf{A (C)} Structure factor of the honeycomb (trigonal) lattice. \textbf{B (D)} analogous cutout from the 928\,eV hydrogen (907\,eV helium) pattern. In \textbf{B} the relative peak intensities are obscured by the strong attenuation of peaks with increasing angle. The relevant features can best be distinguished when comparing peaks with approximately the same distance to the center. Note that the orientation of the pattern is rotated by 90° with respect to the other figures in the manuscript to make the relative intensities in panels \textbf{A} and \textbf{B} more visible.}
    \label{fig:structure_factor}
\end{figure}

For a trigonal lattice, the structure factor is independent of $m$ and $n$.
This is different for a honeycomb lattice, where $\vert F_{nm}\vert^2=4$ (instead of 1) when $m+n$ is a multiple of 3.
The values for the hexagonal and trigonal lattice are visualized together with the observed diffraction patterns in Fig.~\ref{fig:structure_factor}.

As described in the main text, the strong interaction of helium with graphene leads to the region in the center of the hexagon becoming the scattering center instead of the individual atoms. 
These centers are arranged in a trigonal lattice leading to a uniform intensity.

\subsection{Loss of coherence due to phase spread}

Here, we provide another approach to explain the loss of coherence in the phonon simulation, which forms the basis for the data shown in Fig.~\ref{fig:phase_landscape}C and F. 
It is based on the analysis of the phase and its variance due to the thermal motion of the carbon atoms.
First, we consider the general expression for scattering at an arbitrary phase landscape $\phi(x,n)$ where $n$ is the number of phonon snapshots.
The calculation below is performed in 1D but extends trivially to multiple dimensions. 
The intensity $I$ at momentum transfer $k_x$ is given by the incoherent average over the many phonon snapshots
\begin{equation}
    I(k_x) = \left< \left|\int \exp\left[{-i\phi(x,n)}{-i k_xx}\right]\text{d}x\right|^2\right>_n,
\end{equation}
where the phase is given by 
\begin{equation}
\phi(x,n) = \phi_0(x) + \sigma_\phi(x)\varepsilon(x,n).
\end{equation}
Here, $\phi_0(x)$ describes the phase imparted by the static lattice, $\sigma_\phi(x)$ is the standard deviation of the expected oscillations, and $\varepsilon(x,n)$ is a random function, which determines the actual individual realizations (standard deviation $\sigma_\varepsilon = 1$, mean $\mu_\varepsilon = 0$, and the stochastic process governing $\varepsilon$ is stationary). 
This models how the phase changes randomly when the atoms in the crystal lattice oscillate around their equilibrium positions. 
Writing all integrals explicitly results in
\begin{align}
    I(k_x) =& \left< \iint \exp\left[{-i(\phi_0(x)+\sigma_\phi(x)\varepsilon(x,n)-k_xx)}\right]\right.\nonumber\\
    \times&\left.\vphantom{\int} \exp\left[{i(\phi_0(x')+\sigma_\phi(x)\varepsilon(x',n)- k_xx')}\right]\text{d}x\text{d}x'\right>_n.
\end{align}
Since the integration over $x$ and $x'$ happens over a large but finite interval, we can change the order of integration and pull the sample average inside 
\begin{align}
    I(k_x) =& \iint \exp\left[{-i(\phi_0(x)- k_xx)}+{i(\phi_0(x')- k_xx')}\right]\nonumber\\
    \times&\left<\exp\left[{-i(\sigma_\phi(x)\varepsilon(x,n)-\sigma_\phi(x')\varepsilon(x',n))}\right]\right>_n \text{d}x\text{d}x'.
\end{align}
We define 
\begin{equation}
    D = \left<\exp\left[{-i(\sigma_\phi(x)\varepsilon(x,n)-\sigma_\phi(x')\varepsilon(x',n))}\right]\right>_n
\end{equation}
and first continue by further analyzing this term.
In order to carry out this average we need to find a joint probability distribution $p_2(\varepsilon_0,x;\varepsilon_1,x')$, describing the likelihood that in one realization of $\varepsilon(x,n)$ the function takes the value $\varepsilon_0$ at the point $x$ while simultaneously taking the value $\varepsilon_1$ at $x'$.
The solution $p_2$ to this problem needs to have the following properties
\cite{Mandel_Wolf_Coherence_1995}:
\begin{align}
    \int p_2(\varepsilon_0,x;\varepsilon_1,x') \text{d}\varepsilon_0 &= p(\varepsilon_1,x')\\
    \int p_2(\varepsilon_0,x;\varepsilon_1,x') \text{d}\varepsilon_1 &= p(\varepsilon_0,x)
\end{align}
\begin{align}
    \iint \varepsilon_0\varepsilon_1 p_2(\varepsilon_0,x;\varepsilon_1,x') \text{d}\varepsilon_0\text{d}\varepsilon_1 &= \gamma(x-x')\\
    \gamma(0)&=1,
\end{align}
where the single-variable probability distributions $p$ are identical and for the sake of simplicity will be assumed to be Gaussian. 
The correlation function $\gamma$ must only depend on the distance between the two sampled points. 
This is fulfilled by a bivariate Gaussian distribution with $\sigma=1$ and a correlation parameter $\rho$. 
This correlation in the real experiment stems from the fact that the deviation of the phase from the static position is not arbitrarily fine.
Instead, its spatial frequency is limited by the range of the atomic potentials and the distance of the atoms to one another:
\begin{align}
    p_2(\varepsilon_0,x;\varepsilon_1,x') &= \frac{1}{2\pi\sqrt{1-\rho^2}}\\
    &\times\exp\left[\frac{-1}{2(1-\rho^2)}(\varepsilon_0^2-2\rho\varepsilon_0\varepsilon_1+\varepsilon_1^2)\right]\nonumber\\
    \rho &= \gamma(x-x').
\end{align}
Given this distribution, we can replace the average over $n$ with the integral over all $\varepsilon_0$ and $\varepsilon_1$ while considering $p_2$ and calculate $D$ explicitly:
\begin{equation}
    D = \exp\left[-\frac{1}{2}\left(\sigma_\phi(x)^2-2\gamma(x-x')\sigma_\phi(x)\sigma_\phi(x')+\sigma_\phi(x')^2\right)\right].
\end{equation}
Re-inserting this into the diffraction integral yields
\begin{align}
    I(k_x) =& \iint \exp\left[{-i(\phi_0(x)- k_xx)}-\frac{1}{2}\sigma_\phi(x)^2\right]\nonumber\\
    &\times\exp\left[{i(\phi_0(x')- k_xx')}-\frac{1}{2}\sigma_\phi(x')^2\right] \nonumber\\
    &\times\exp{\left[\gamma(x-x')\sigma_\phi(x)\sigma_\phi(x')\right]}\text{d}x\text{d}x'.
\end{align}
If the correlation is $\gamma(x-x')\equiv0$, this separates out into a diffraction integral where the phase spread merely contributes as an amplitude reduction of $\exp(-\sigma_\phi(x)^2/2)$:
\begin{align}
    I(k_x)_{\gamma\equiv0} &= \left|\int \exp\left[{-i(\phi_0(x)- k_xx)}-\frac{\sigma_\phi(x)^2}{2}\right] \text{d}x\right|^2.
\end{align}
Realistically, the correlation is not completely zero, but extends for some range $a$ before approaching zero. However, if our integration interval (i.e. our scattering element's size) is large with respect to the correlation range $a$, then the approximation of $I(k_x) \approx I(k_x)_{\gamma\equiv0}$ gets better and better. For the experiment this means that transverse coherence must be large with respect to the range of the potential/atomic bond length. 

\subsection{Comparison to phase accumulated by electrons in TEM}

\begin{figure}[b]
    \centering
    \includegraphics[width=\linewidth]{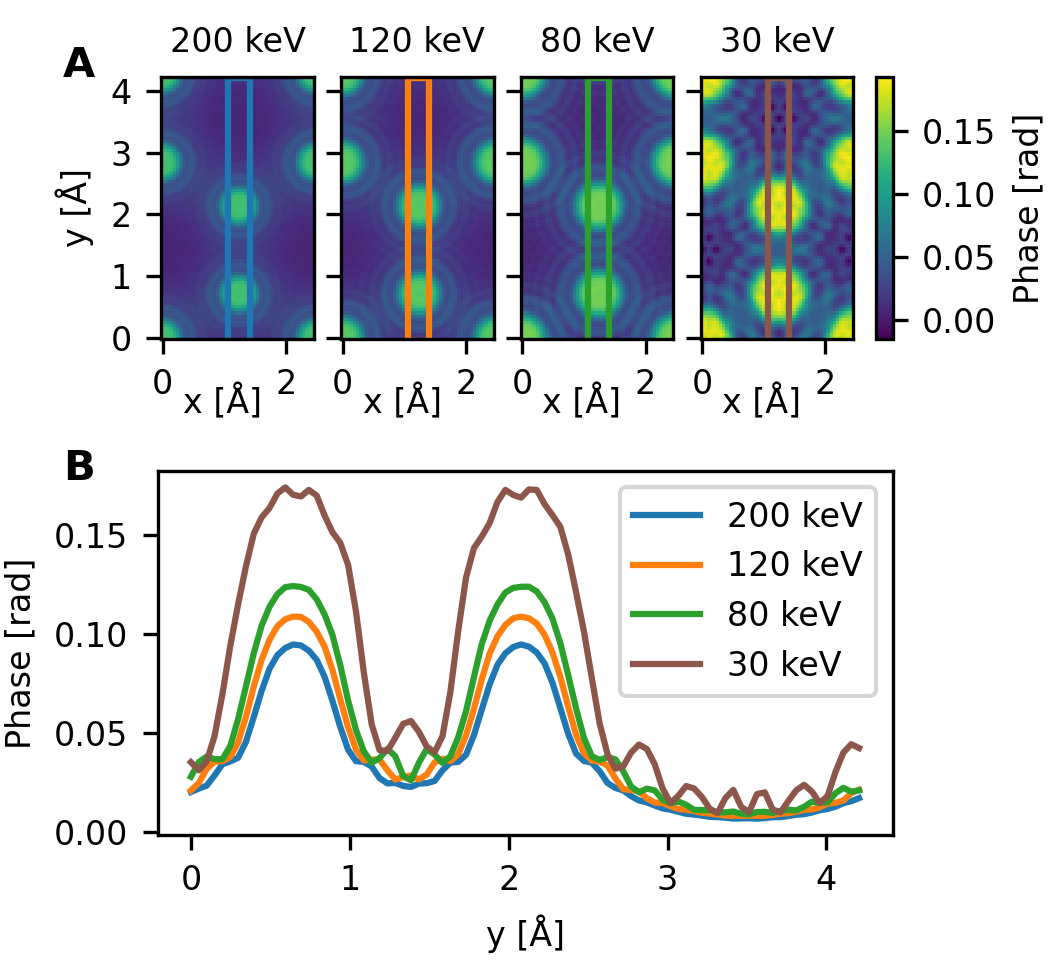}
    \caption{Simulated accumulated phase of electrons at energies typical for transmission electron microscopy. \textbf{A} Phase images of the complex exit waves. \textbf{B} Vertical (bottom to top) line profiles over the colored rectangles in \textbf{A}.}
    \label{fig:tem}
\end{figure}

To compare the magnitudes of the accumulated atomic phases to those of electrons at energies typical for transmission electron microscopy (TEM), we simulated the complex exit-waves for electron plane-waves transmitting through single-layer graphene at energies between 30 and 200~keV using the standard multislice method as implemented in the \emph{ab}TEM code~\cite{madsen_abtem_2021}. The scattering potentials were generated using the independent atom model with an $xy$-sampling of 0.05~\AA\ and a single slice in the $z$-direction perpendicular to the lattice. Figure~\ref{fig:tem} plots the phases of the complex exit-waves and corresponding line profiles. For electron energies of (200, 120, 80, 30)~keV, the maximum phase shifts over the atoms are (0.095, 0.109, 0.124, 0.174)~rad, while the minimum values over the hexagon centers are (0.007, 0.008, 0.009, 0.010)~rad. We thus conclude that the atomic phases shown in Fig.~\ref{fig:phase_landscape} are two orders of magnitude higher than typically accumulated by electrons in TEM.

\subsection{Additional traces through diffraction images}

\begin{figure*}
    \centering
    \includegraphics[width=1\linewidth]{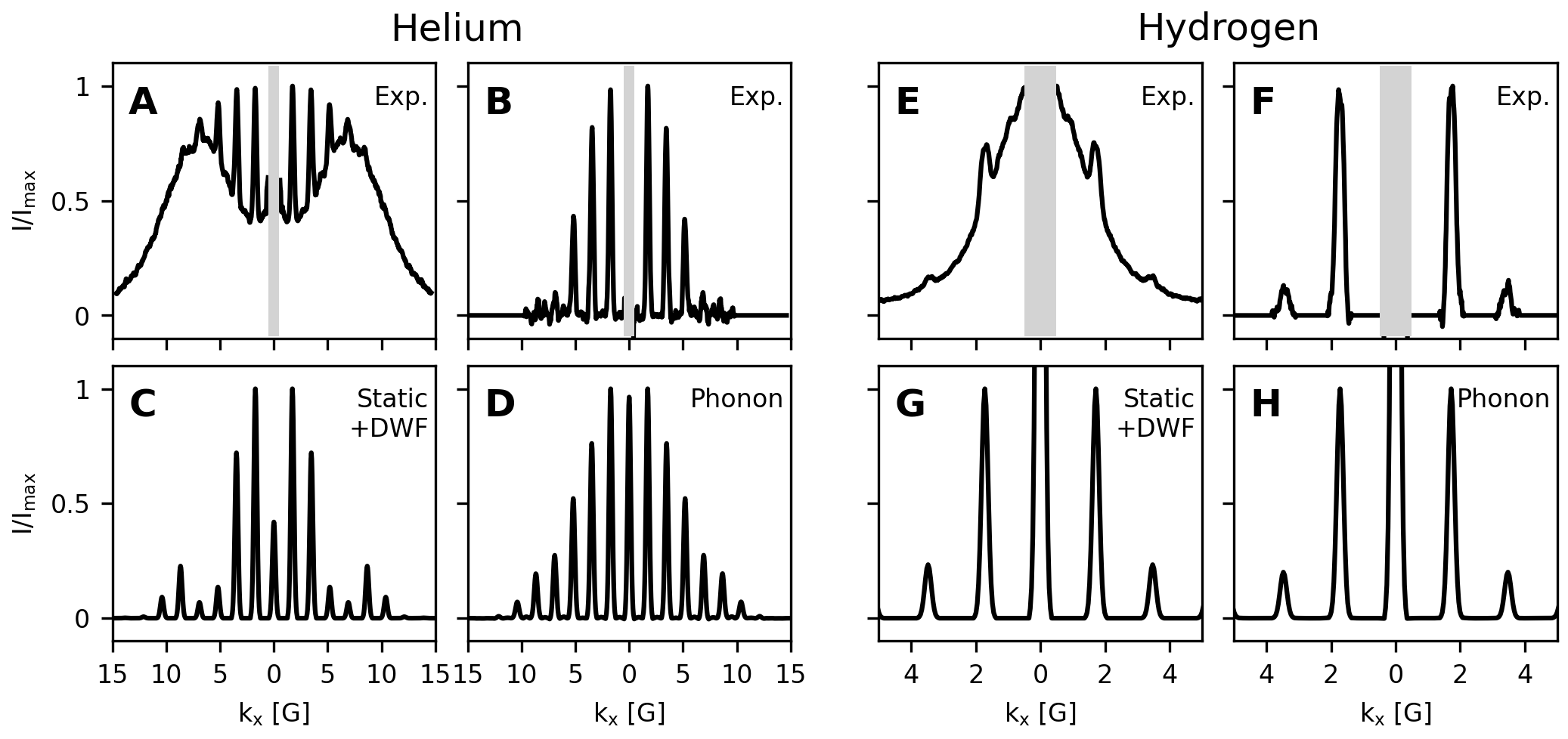}
    \caption{Traces along the $k_x$ direction for the data shown in Fig.\,\ref{fig:fig_He} and Fig.\,\ref{fig:fig_H}. \textbf{A-D} shows helium while \textbf{E-H} shows hydrogen. \textbf{A (E)} Horizontal slice through the full diffraction pattern and \textbf{B (F)} the background-subtracted trace. \textbf{C (E)} Simulation of the static lattice with applied DWF. \textbf{D (H)} Background-subtracted phonon simulation.}
    \label{fig:other_direction_traces}
\end{figure*}

In the main text, the one-dimensional cuts through the diffracted pattern along the $k_y$ direction are shown. 
The traces along the orthogonal $k_x$ direction are presented in Fig.~\ref{fig:other_direction_traces}. 
To ease the comparison, they are scaled to the maximum intensity in the experimental pattern.

\subsection{Agreement of different simulation methods: Hydrogen}

\begin{figure*}
    \centering
    \includegraphics[width=\linewidth]{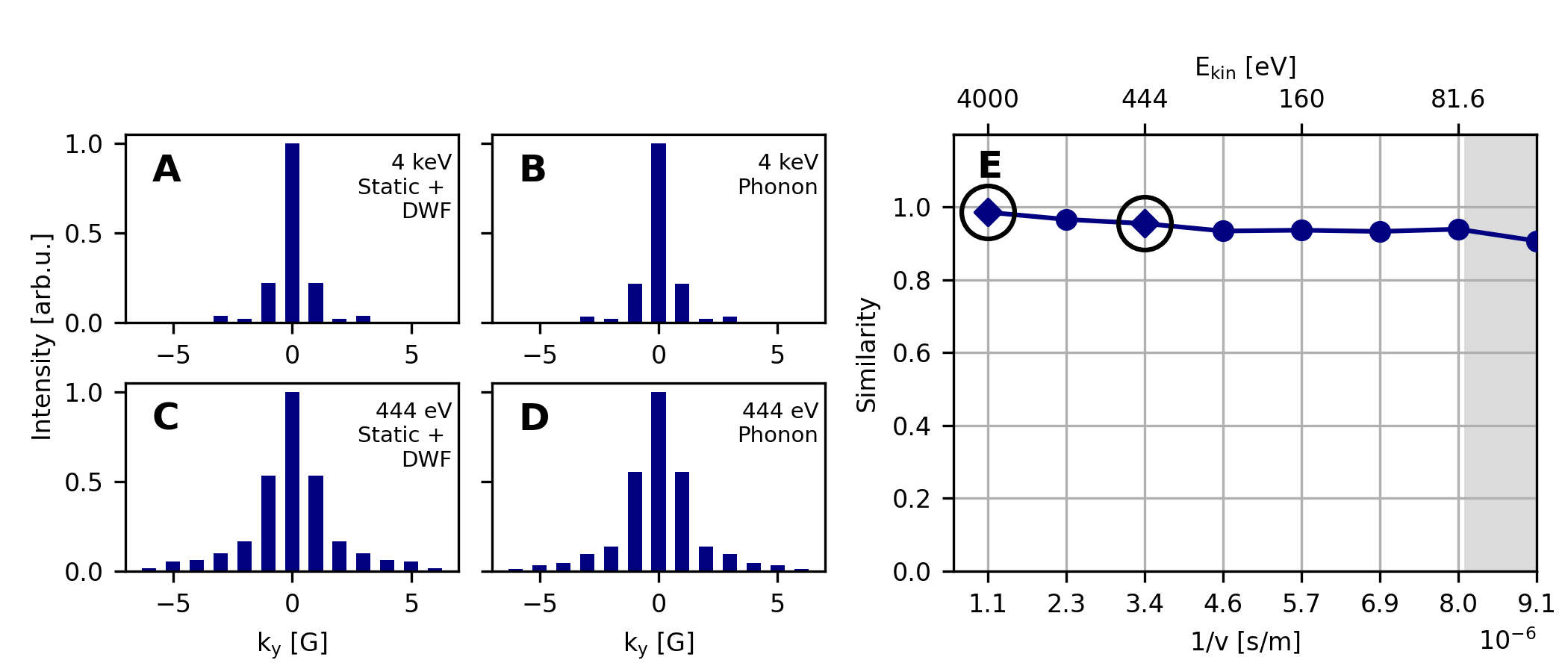}
    \caption{\textbf{Simulations of diffraction peak intensities at varying energies using two different simulation methods for hydrogen}. (\textbf{A} and \textbf{C}) Peak intensities for hydrogen diffraction at 4\,keV  and 444\,eV generated from the static lattice geometry and modulated by a DWF. (\textbf{B} and \textbf{D}) Hydrogen simulation at 4\,keV and 444\,eV averaging over the diffraction images of 100 phonon-distorted lattice configurations.
    (\textbf{G}) Similarity score $S$ of the phonon and DW simulations at varying kinetic energies and plotted as a function of $1/v$.
    The circled data points designate the data shown in panels (\textbf{A}-\textbf{D}). 
    The grey area designates the breakdown of the eikonal approximation.}
    \label{fig:simulation_compare_h}
\end{figure*}

In Fig.~\ref{fig:simulation_compare} of the main text, the energy-dependent agreement between the two applied simulation methods  for helium is shown. 
Figure~\ref{fig:simulation_compare_h} shows the corresponding comparison for hydrogen, using the same values for $1/v$. 
The similarity score is above 0.9 at all considered energies and even at low energies the showcased diffraction patters are very similar.

\subsection{Background-subtraction of simulations}

\begin{figure*}
    \centering
    \includegraphics[width=1\linewidth]{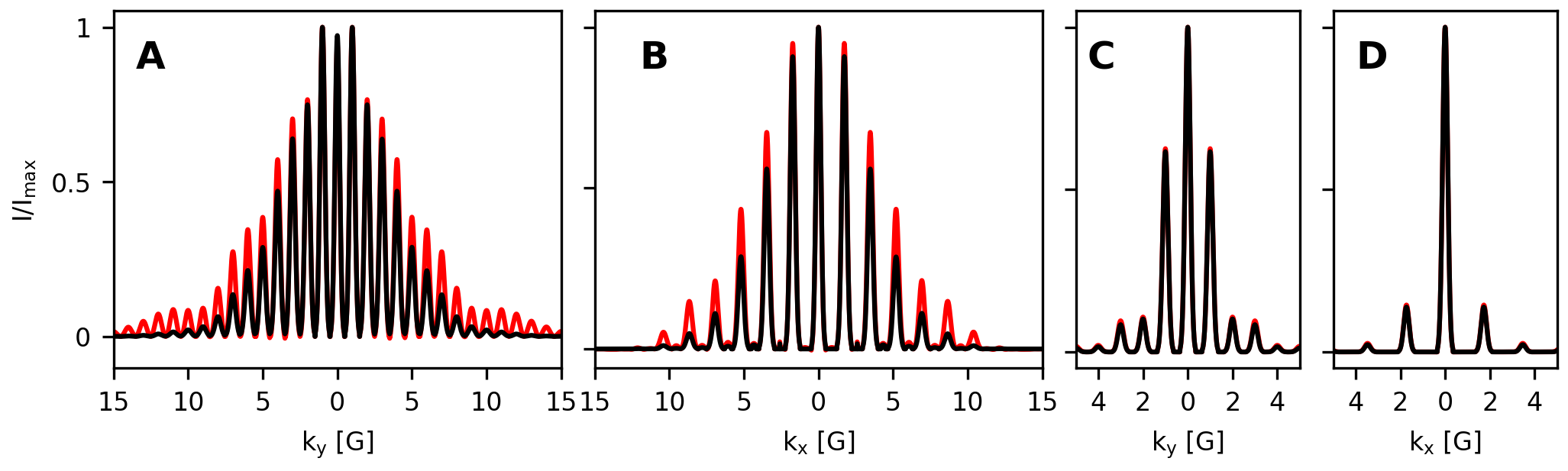}
    \caption{Comparison traces for different approaches of determining the diffraction peaks above the background in the phonon simulations. Red: diffraction signal after background determination by interpolation; black: diffraction signal after background selection by rejecting all intensity that is not at the exact theoretical position before collimation is applied. \textbf{A} (\textbf{B}) Helium at 907 eV along the $k_y$ ($k_x$) direction. \textbf{C} (\textbf{D}) Hydrogen at 928 eV along the $k_y$ ($k_x$) direction. For hydrogen the red traces are almost entirely obscured by the back ones because the results are nearly identical.}
    \label{fig:background_subtracted}
\end{figure*}

\begin{figure*}
    \centering
    \includegraphics[width=1\linewidth]{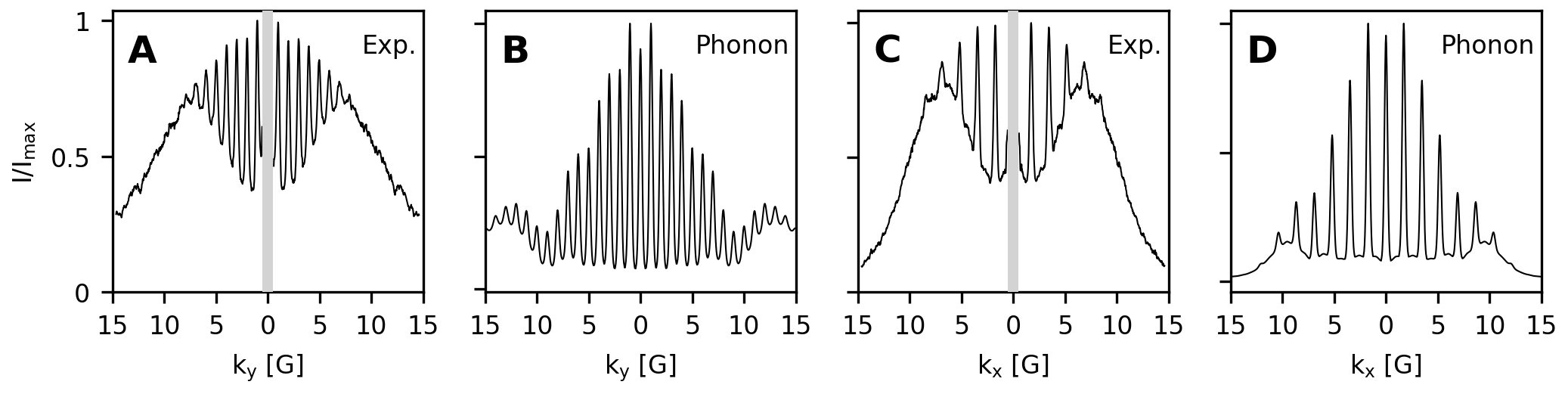}
    \caption{Comparison between helium diffraction data and simulation without background subtraction.  \textbf{A} (\textbf{C}) Vertical (horizontal) profile through the experimental diffraction pattern shown in Fig.~\ref{fig:fig_He}A. \textbf{B} (\textbf{D}) analogous profiles through the full phonon simulation.}
    \label{fig:background_not_subtracted}
\end{figure*}

The phonon simulations shown in Fig.~\ref{fig:fig_He}E yield both peaks and background contributions. 
There are multiple ways of discriminating between the foreground and the background for this data. 
We have chosen to treat the simulations comparable to the recorded data and determine the background by interpolation between the diffraction peaks. 
Alternatively, one can select the contribution at the theoretically expected positions as the foreground before collimation is applied to the image. 
In Fig.~\ref{fig:background_subtracted} we compare these two approaches. It can be seen that for helium the background determination via interpolation (red traces) discards less of the signal as background compared to the version that only selects the value at the exact theoretical positions (black traces).
The background created in the hydrogen simulations is very weak and can be neglected.
Figure~\ref{fig:background_not_subtracted} shows the helium simulations without background subtraction.

\subsection{Validity of the eikonal approximation}

\begin{figure*}
    \centering
    \includegraphics[width=1\linewidth]{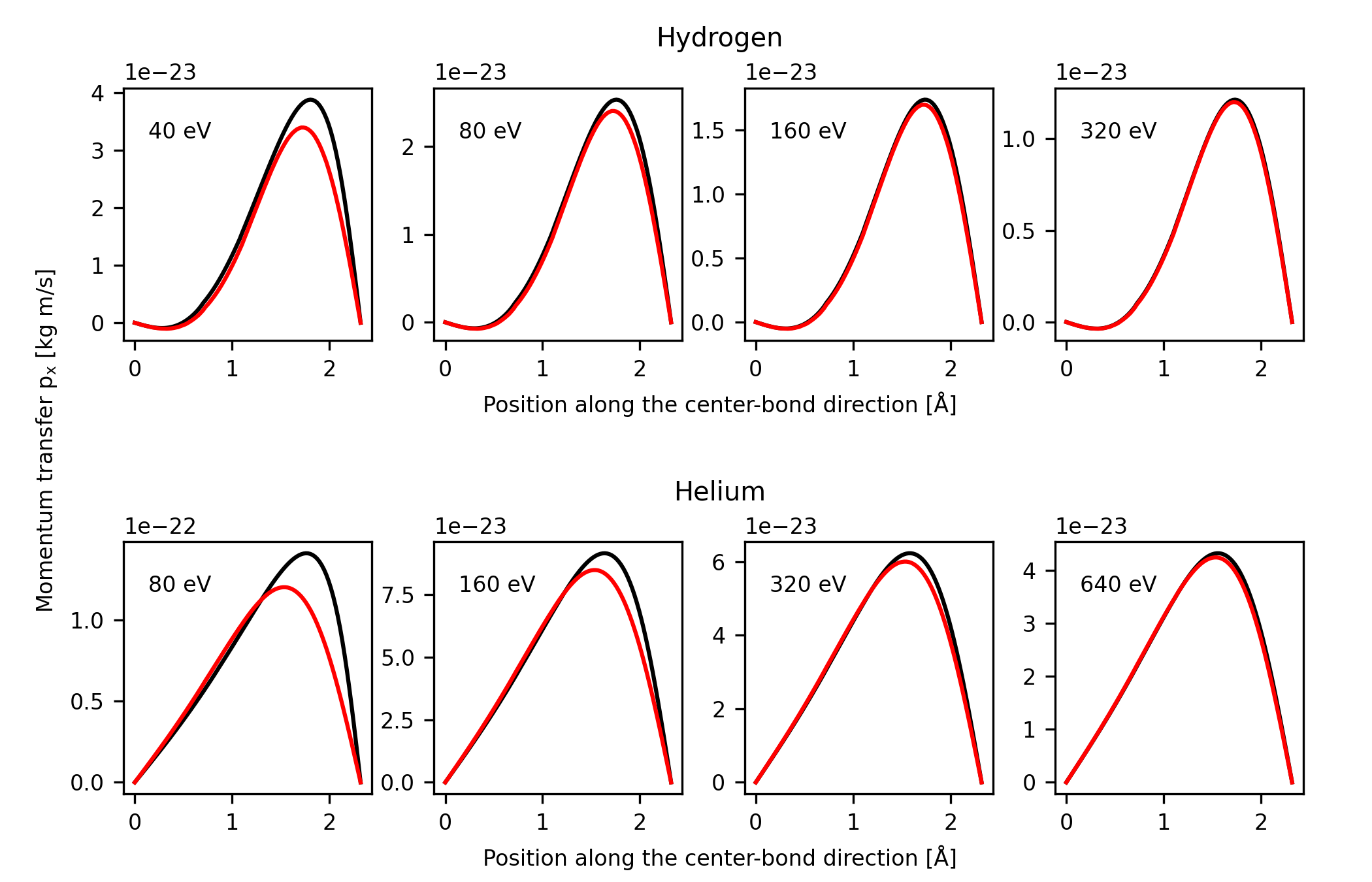}
    \caption{Classical deflection simulations for hydrogen and helium using two different methods. Black traces: Deflection according to the 3D Hamilton equations; Red traces: Deflection according to the eikonal approximation. Deflections were calculated along the line connecting the center of a hexagon to the nearest bond.}
    \label{fig:Eikonal_Validity}
\end{figure*}

In Fig.~\ref{fig:simulation_compare} we compare different ways to model the influence of lattice distortions over a large range of energies. 
Both approaches are done within the eikonal approximation and, as the kinetic energy takes on lower values, this approximation  breaks down at some point. 
In order to determine at which point that occurs, an additional set of calculations was performed. 
Fig.~\ref{fig:Eikonal_Validity} shows the classical deflection calculated two different ways for different energies for both helium and hydrogen. 
The red traces show the horizontally imparted momentum according to the eikonal approximation while the black traces show the imparted momentum when calculating the trajectories using the 3D Hamilton equations. 
The hydrogen simulations are very similar upwards of 80\,eV, which is in agreement with earlier reports~\cite{Brand_NewJPhys21_033004}. 
For helium, the simulations converge for energies above 320\,eV. 
The region where the eikonal is not valid is marked with a light grey background in Figs.~\ref{fig:simulation_compare} and \ref{fig:simulation_compare_h}.

\end{document}